\documentclass[twocolumn]{svjour3}          

\usepackage{times}
\usepackage{graphicx}%
\usepackage{multirow}%
\usepackage{amssymb,amsfonts}%
\usepackage{mathrsfs}%
\usepackage[title]{appendix}%
\usepackage{booktabs}%
\usepackage{siunitx}
\usepackage{amsmath}
\usepackage{orcidlink}
\usepackage{mathtools}
\usepackage{bm}
\usepackage{multicol,lipsum}
\usepackage{cuted}
\usepackage{titlesec}
\usepackage{orcidlink}
\usepackage[normalem]{ulem}
\usepackage{cancel}
\usepackage[]{xcolor}
\usepackage[final]{changes} 
\definechangesauthor[name=new additions, color=blue]{new}
\definechangesauthor[name=in place modifications, color=blue]{modified}
\definechangesauthor[name=deletions, color=red]{deleted}

\titlespacing*{\section}{0pt}{0.5\baselineskip}{0.2\baselineskip}
\titlespacing*{\subsection}{0pt}{0.5\baselineskip}{0.2\baselineskip}
\titlespacing*{\subsubsection}{0pt}{0.5\baselineskip}{0.2\baselineskip}
\hypersetup{hidelinks} 

\begin{document}

\title{Motor State Prediction and Friction Compensation for Brushless DC Motor Drives Using Data-Driven Techniques}

\author{Nimantha Dasanayake \orcidlink{0000-0001-6221-1610} \and Shehara Perera \orcidlink{0000-0001-6695-4307}}

\institute{Nimantha Dassanayake \at
              Department of Mechanical Engineering, University of Moratuwa, Katubedda 10400, Sri Lanka \\
              \email{dasanayakenp.21@uom.lk} 
           \and
           Shehara Perera \at
              Department of Engineering, University of Cambridge, Cambridge CB3 0FA, UK\\
              \email{ulsp2@cam.ac.uk}
}

\date{Received: DD-MM-2024/ Accepted: DD-MM-2024}

\maketitle

\begin{abstract}
Friction compensation is critical for robust, dependable, and accurate position and velocity control of motor drives. Large position inaccuracies and vibrations caused by non-characterised friction may be amplified by stick-slip motion and limit cycles. This research study uses two data-driven methodologies to find the governing equations of motor dynamics, which also describe friction. Specifically, data obtained from a brushless DC (BLDC) motor is subjected to the Sparse Identification of Nonlinear Dynamics with Control (SINDYc) technique and low-energy data extraction from time-delayed motor velocity coordinates to determine the underlying dynamics. Next, the identified nonlinear model was compared to a linear model without friction and a nonlinear model that contained the LuGre friction model. The optimal friction parameters for the LuGre model were determined using a nonlinear grey box model estimation approach with the collected data. The three validation datasets taken from the BLDC motor were then used to validate the resultant innovative nonlinear motor model with friction characteristics. Over $90\%$ accuracy in predicting the motor states in all input excitation signals under consideration was demonstrated by the innovative model. Additionally, when compared to a system that was identical but used the LuGre model, a model-based feedback friction compensation technique demonstrated a relative improvement in performance.

\keywords{System identification \and SINDYc \and Friction characterisation \and Motor control}

\end{abstract}

\section{Introduction}
    \label{Introduction}

Friction identification and compensation are critical components of actuator control to avoid substantial positional inaccuracies, stick-slip effects, and limit cycles, resulting in smooth and accurate trajectories \cite{[1],[4]}. The highly nonlinear nature of rotor dynamics, primarily due to friction, makes it challenging to find a suitable mathematical model for developing high-performance controllers. The typical friction components identified in experimental data are Coloumb, Stribeck, and Viscous. Improvements in friction modelling and parameterization research revealed that pre-sliding friction and sliding friction comprise the friction regime. Researchers also mention an intermediate phase called transition friction in the literature \cite{[2],[3],[4]}. Friction compensation has piqued the interest of researchers in a wide range of control engineering subfields. Robotics, autonomous systems, and vehicle dynamics are just a few of the many fields that could greatly benefit from friction compensation, gaining dependable and resilient control over their intrinsically nonlinear systems.

\subsection{Friction Models}
    \label{Friction_Models}

\begin{figure*}[tbp]
    \centering
    \input{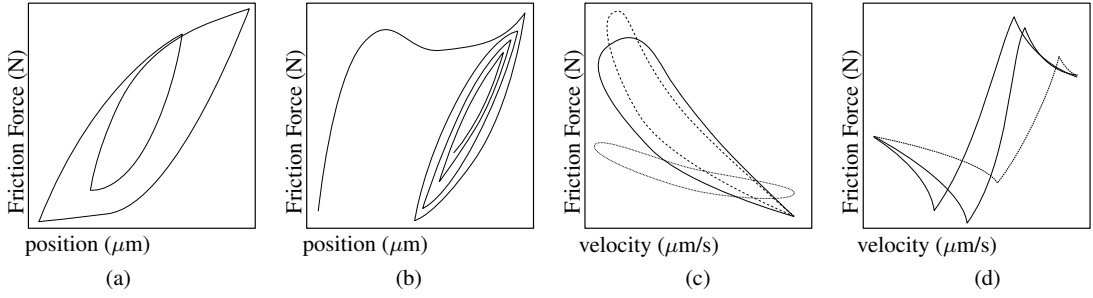}
    \vspace{-5pt}
    \caption{The four main properties of friction (a) hysteresis, (b) frictional lag, (c) periodic friction and (d) non-drifting property}
    \vspace{-10pt}
    \label{Properties}
\end{figure*}

Asperities at the microscopical level cause the frictional forces between two contacting surfaces, according to study results. The effects of these asperities are dependent on factors like displacement and relative velocity of the surfaces, the presence of lubrication, normal forces, and temperature \cite{[1]}. The complex nature of the interaction between surfaces gives rise to more than one regime of frictional behaviour: pre-sliding, transition, and gross sliding \cite{[12]}.

Different models that capture the friction components at different levels have previously been proposed \cite{[1-2]}. These models can be basically categorised as static or dynamic.
The static models' weakness is their inability to capture the forces in a pre-sliding friction regime. Further studies have revealed that a hysteretic displacement-dependent friction force is present in the pre-sliding region \cite{[12]}. A successful friction model must incorporate three other properties, namely frictional lag, periodic friction, and non-drifting properties, in addition to hysteresis. In \cite{[18]}, a comparison of these characteristics for different friction models has been presented.

The first dynamic friction model that was developed is called the Dahl model. This model not only captures the pre-sliding behaviour but also the hysteresis effect \cite{[12]}. The Dahl model first characterised pre-sliding friction's position dependence and velocity independence (rate independence). If the external force is not enough to exceed static friction, the asperities at the contact surface will deform, resulting in a pre-sliding motion \cite{[1-3]}. The force in the Dahl model is characterised by the following functions:
\begin{gather}
F_f =   kz \label{eq:10}\\
\frac{\mathrm{d}z}{\mathrm{d}t} = v\bigg(1 - \mathrm{sgn}(v)\frac{\sigma_z}{F_c} \bigg) \label{eq:11}
\end{gather}
\noindent \added[id=new]{Here, $F_f$ is the total frictional force, $F_c$ is the Coulomb friction, $z$ is the asperity deformation, $k$ is the stiffness of the asperities and $v$ is the velocity between sliding surfaces.} However, as shown in Equations \ref{eq:10} and \ref{eq:11}, the Dahl model does not capture stiction or Stribeck friction, which is rate-dependent. Later on, many enhanced models have developed to capture these effects. One such model is the popular LuGre model, in which the asperities are modelled as small bristles \cite{[3]}. The equations are as follows:
\begin{gather}
\mathit{F_f =   kz + \sigma \frac{\mathrm{d}z}{\mathrm{d}t} + \alpha_2 v} \label{eq:12}\\
\mathit{\frac{\mathrm{d}z}{\mathrm{d}t} = v\bigg(1 - \mathrm{sgn}(v)\frac{kz}{s(v)} \bigg)} \label{eq:13}\\
\mathit{s(v) = \alpha_0 + \alpha_1 e^{-{\big(\frac{v}{v_s}\big)}^2 }} \label{eq:14}
\end{gather}

\noindent Here, \added[id=new]{$\alpha_0 = F_c$ and $\alpha_1 = F_s - F_c$ where $F_s$ is the Strikbeck friction. The Stribeck velocity is denoted by $v_s$. The coefficient $\sigma$ gives the asperity deformation rate dependency to the overall frictional force.} Furthermore, $\alpha_2$ is the viscous friction coefficient. It is evident that the LuGre model has been used successfully in many applications in developing control systems with friction compensation \cite{[27],[28],[29]}.

The non-differentiable point at breakaway velocity represents one of the LuGre model's drawbacks. To overcome this issue, a new continuously differentiable friction model has been proposed in \cite{[7]} and applied in \cite{[4]}. Furthermore, the LuGre model lacks the non-drifting property (shown in Figure \ref{Properties}(D), which was later extended using an elastoplastic model of bristles instead of the pure elastic model \cite{[11]}. However, the non-local memory hysteresis behaviour of friction is not characterised by Dahl or LuGre models. Attempts have been made to modify the model so that this property is captured. For instance, the Leuven model described in \cite{[26],[22]} introduces a hysteresis function ($F_h$) to the LuGre model. However, due to the complexity of the model, it is difficult to implement. \added[id=new]{The disagreement of LuGre and Leuven models with some experimental results has motivated the researchers to develop a model that replaces the hysteresis function with a Maxwell Slip model representation called the Generalised Maxwell Slip (GMS) model \cite{[18]}.}

Several studies have been carried out to compare the friction models discussed so far. In \cite{[3]}, GMS models with 4 and 10 elements were compared with the LuGre model, and it was shown that GMS models outperform LuGre. In \cite{[11]}, a GMS model of up to 10 blocks has been tested and concluded that at least 4 blocks are essential to accurately modeling the friction. \cite{[31]} shows that the GMS model performs better than LuGre. This proves that GMS captures most of the friction properties and gives better predictions when compared with other models. But for the downside, it must also be mentioned that to identify the parameters, high-resolution position measurement is needed. As a result, it was not possible to fit a GMS model during the work carried out for this research study.

\subsection{Friction Compensation}
    \label{Friction_Compensation}

 \begin{figure}[tbp]
 \begin{center}
\includegraphics[width=0.5\textwidth]{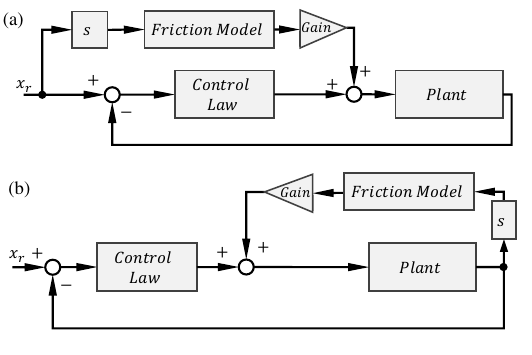}
 \vspace{-15pt}
 \caption{Common friction compensation algorithms, (a) Feedforward, (b) Feedback}
 \vspace{-25pt}
 \label{fig:Feedback}
 \end{center}
 \end{figure}

Several approaches have been proposed to compensate for friction in motor drives. These approaches consider different types of friction phenomena, use various models to represent them, and employ different strategies to reduce or eliminate their adverse effects on motion. This section will provide concise overviews of four categories of friction compensation: feedforward and feedback compensation, compensation in variable structure systems, observer-based compensation, and adaptive friction model-based compensation.

\subsubsection{Feedforward and Feedback Compensation}\
Feedforward compensation in various studies has utilised reference inputs, such as position or velocity, in order to predict future force values \cite{[2],[3],[31]}. A compensation scheme has been proposed in \cite{[2]} and \cite{[3]}. This scheme estimates the inertia and friction forces using feedforward reference position and modifies the control signal from a Proportional Derivative (PD) control. This method has resulted in a significant reduction of over $80\%$ in positional error compared to the control without compensation. The feedforward term in the study referenced by \cite{[15]} includes a component that takes into consideration the velocity error. Unlike feedforward compensation, the feedback compensation technique utilises the measured outputs, such as position and velocity, to estimate the friction force. Then it is fed back to the control signal with a gain. The majority of systems developed to date have utilised the LuGre model for this estimation \cite{[33],[35]}. A comparison between the two approaches can be done using the block diagrams shown in Figure \ref{fig:Feedback}.

\subsubsection{Compensation in Variable Structure Systems}

Variable Structure Systems (VSS) are systems that adapt the control model structure to account for nonlinearities. VSS are suitable for controlling systems that involve nonlinear friction forces. As an example, researchers have utilised sliding mode control (SMC) in various studies to effectively achieve accurate position and/or velocity tracking \cite{[27],[28],[36],[4]}. A study has been conducted on the design of an adaptive robust controller with a continuous friction model, as well as the use of an adaptive funnel sliding mode control in another study. The authors of \cite{[40]} explore a unique perspective on friction. They discuss situations where friction serves as a restoring force, bringing the dynamics back to the sliding surface without compensation. Another example of a VSS is the use of robust integral of the sign of the error (RISE) feedback term to design an innovative adaptive controller to compensate for nonlinear friction and bounded disturbances \cite{[39]}.

\subsubsection{Observer based Compensation}

Observers have been used to estimate the internal variable (asperity deformation), thus estimating the friction \cite{[5],[1-2]}.
A non-model-based extended state reduced order observer (ESO)  with only first order derivatives has been developed for robot joint motion in \cite{[32]}. Another ESO-based control approach can be found in \cite{[34]}. Furthermore, the Extended Kalman-Bucy Filter (EKBF) has been used together with a dynamic friction model to observe the friction force under varying normal load conditions \cite{[5]}. The system is adaptive to normal load changes and they have tried both LuGre and Dahl models. Around $75\%$ decrease in error when compared with PID control without friction compensation can be seen in the results. In addition, observers based on LuGre, Dahl and Bliman-Sorine models are discussed in \cite{[1-2]}.

\subsubsection{Adaptive Friction Model based Compensation}

Although friction has been modelled accurately enough to improve the position and velocity tracking of systems, unless these models are adaptive, they can give inaccurate friction estimates due to the uncertainty of model parameters. For instance, the adaptive compensation strategy discussed in \cite{[3]} has considered the Coulomb friction and Viscous friction as the adaptive parameters of the LuGre model since these have the most significant effect on the accuracy of friction estimation. The two parameters have been multiplied by two gains, which are updated according to a law that accounts for the error in the tracked position. In \cite{[27]} a similar formulation, an adaptive LuGre model has been used to compensate friction in an adaptive SMC scheme known as adaptive non-singular fast terminal SMC. Furthermore, an adaptive robust control implemented for position tracking using a novel continuous friction model can be seen in \cite{[4]}. 

\added[id=new]{
The major contribution of this research study is threefold:}
\begin{itemize}
    \item \added[id=new]{Firstly, it demonstrates that data-driven techniques can effectively extract the internal state of friction from velocity data. The extraction method is proposed by authors by uniquely combining time delay embedding and SINDYc algorithm.}
    \item \added[id=new]{Secondly, it demonstrates that a motor model constructed using data-driven nonlinear system identification techniques exhibits greater accuracy against one of the most commonly used friction models found in the literature. Moreover, the model also captures the hysteresis with non-local memory, a characteristic that is challenging to replicate using current models.}
    \item \added[id=new]{Thirdly, it shows the efficacy of the identified motor model for feedback friction compensation. A comparison of friction compensation performance with a commonly used friction model validates the advantage of the proposed model.}
\end{itemize}


The rest of the article is structured as follows: Section \ref{System_Identification} provides a discussion of techniques for identifying nonlinear systems. Next, the system identification and friction compensation strategies employed in this study are detailed in Section \ref{Mathematical Modelling}. The experiment setup and methodology are described in Section \ref{Experiments}. The results obtained after adopting the proposed methodology are reported in Section \ref{Results}. The article is concluded in Section \ref{Conclusion} with a description of the future directives.

\vspace{5pt}

\section{System Identification}
    \label{System_Identification}
Finding governing equations from data is a significant challenge for nonlinear systems \cite{[43]}. Mechanical systems that experience friction are difficult to regulate without a nonlinear dynamics model. A key contribution of this research is the unique application of \added[id=new]{two established system identification methods for friction model discovery in an electric motor}. This section will introduce the Sparse Identification of Nonlinear Dynamics with Control (SINDYc) technique (\cite{[44]}). The theory of Time Delay Embedding (TDE) and Dynamic Mode Decomposition (DMD) will be explained, along with Koopman theory insights (\cite{[54]}). Although the authors' DMD model is not presented in this article due to its low performance, the theory is presented for completeness of the literature review and since it is the motivation behind the extraction of the internal state of friction.

\subsection{Sparse Identification of Nonlinear Dynamics with Control (SINDYc)}
Sparse Identification of Nonlinear Dynamics (SINDy) is an algorithm designed to extract parsimonious dynamics from time-series data using a regression method. Models for fluid flows, optical systems, chemical reactions, plasma convection, structures, and model predictive control have all been identified through the extensive use of SINDy.  \cite{[45]}. For most of the physical systems, it can be assumed that the structure of the governing equation only contains a few important terms so that the equations are sparse in the space of possible functions \cite{[43]}. Under this assumption, sparsity-promoting techniques in cooperation with machine learning have been used to discover nonlinear dynamics through the regression of data. For example, \cite{[43]} presents the results of predicting the dynamics of a simulated chaotic Lorenz system and vortex shedding of a flow passing a cylinder with high accuracy. The SINDy algorithm has been generalized to systems with external control inputs: SINDYc \cite{[44]}. An advantage of this method is balancing model complexity with descriptive ability, thus promoting interpretability and generalizability \cite{[45]}.

\subsection{Time Delay Embedding and Dynamic Mode Decomposition}

Multiple research studies have focused on determining the appropriate coordinate transformations for measuring dynamical systems to approximate nonlinear dynamics with a higher-order linear system \cite{[53],[52],[45]}. The Koopman Theory is the motivation behind using linear systems to approximate nonlinear dynamics \cite{[54]}. A finite-dimensional nonlinear dynamical system can be represented by an infinite-dimensional system made of observable functions of Hilbert space states, according to the theory. Koopman theory has been used commonly in fluid dynamics \cite{[55]} and discovering dynamics of chaotic systems \cite{[52]}. 

In recent developments in the field of data-driven model identification, various techniques have been tested to find an approximated finite dimensional subspace of the Koopman observables, so the theory can be practically implemented to predict the behaviour of nonlinear dynamical systems. For high dimensional systems this has been done using a method called Hankel Alternate View of Koopman (HAVOK) that is based on TDE and DMD \cite{[52]}. HAVOK has been applied to predict the dynamics of chaotic systems like the Lorenz attractor. It has also been applied to identify the unknown dynamics of soft robots \cite{[47]}.

When the full state of the model is not measurable due to the existence of hidden variables, or lack of measured data, it is possible to reconstruct the dynamics of the system approximately using a higher dimensional model that has time-delayed versions of the available state measurements as new states \cite{[53]}. By singular value decomposition (SVD) of the delay coordinates and DMD of the Eigenmode time series data, a promising low dimensional approximation to a Koopman system could be obtained \cite{[52]}. DMD can also be done to forced nonlinear systems and the method is comprehensively discussed in \cite{[46]}.

The authors were driven by the possibility of reconstructing inaccessible or internal system states using data from only one or a few states. As a result, a hypothesis was formed: the internal state of friction (specifically, asperity deformation as explained in Section \ref{Friction_Models}) can be reconstructed using velocity data by selecting an appropriate order for TDE and extracting components that are significant for low energy singular values of SVD. Further details of the followed methodology are presented in Section \ref{Mathematical Modelling}.

\section{Mathematical Modelling}
    \label{Mathematical Modelling}
It was ascertained through the study of state-of-the-art that fewer attempts have been put \added[id=new]{into} incorporating the data-driven methods to build motor models that account for frictional forces. Therefore, the main objective of this research was to discover how effective the data-driven methods are in replacing the conventional friction models to give better predictions on motor states. Specifically, the methodology followed in this research can be described in two steps: first, TDE was used to build up a high-order coordinate system from motor velocity data to extract the internal state of friction. As the second step, a data-driven nonlinear model identification method: SINDYc was used to fit a motor model incorporating internal state data extracted from the previous step. All data for the experiments were obtained by exciting a BLDC motor using an excitation signal that is described in subsequent sub-sections. Furthermore, three more excitation signals were devised to validate the models to see their ability to be generalized for a wider span of operating conditions. The data-driven method was then compared with the conventional linear motor model and a motor model with nonlinear friction. Considering all the friction models discussed in Section \ref{Friction_Models}, it was decided to use the LuGre model considering its large number of practical implementations found in literature and the feasibility of determining the parameters correctly using a low-resolution position measurement. 

For a rigorous treatment for the proof of the concept presented in this article, it was necessary to compare three dynamic models: the first two models are existing ones and the third one is a novel model based on data-driven techniques, that was developed by authors. 

\subsection{Linear Model with Viscous Friction}

The first model was the conventional linear motor dynamic model that considers only viscous friction which is described in Equations \ref{eq:1} and \ref{eq:2}. The coefficients for the model were identified using linear regression. The method is not described here since it is insignificant to the focus of this article. 

\subsection{Nonlinear Model with LuGre Friction}

The second model was obtained by modifying the first model by introducing the LuGre friction model described by Equations \ref{eq:12} - \ref{eq:14}. The parameter determination method used for this model was adopted from \cite{[3]}. The static parameters of the LuGre model i.e $\alpha_0$, $\alpha_1$, $\alpha_2$ and $v_0$ were determined using steady-state velocity responses of the motor to low excitation voltages. The initial values of the dynamic parameters i.e. $\sigma_0$ and $\sigma_1$ were calculated based on the current responses for slow ramp excitation voltages. The final values of dynamic parameters were identified using the estimation of a nonlinear grey box model. The search method used for the estimation was the Trust-Region Reflective Newton method of nonlinear least-squares. This is a constrained optimisation method explained by the following equation,
\begin{equation}
\displaystyle{\min_{\bm{\xi}} \added[id=new]{G}(\bm{\xi})} = \displaystyle{\min_{\bm{x}}{||\bm{y-f(x,u,\xi)}||^2_2}}
\label{eq:26}
\end{equation}
\noindent where, $\added[id=new]{G}$ is the nonlinear cost function, $\bm{x} \in \mathbb{R}^{n}$ is the state vector, $\bm{y = \dot{x} \in \mathbb{R}^{n}}$ is the derivative vector, $\bm{u} \in \mathbb{R}^{m}$  is the input vector, $\bm{\xi} \in \mathbb{R}^{l}$ is the vector of unknown coefficients and  $\bm{f:\mathbb{R}^{n}  \rightarrow \mathbb{R}^{n}}$ is a continuous function that defines the time evolution of the dynamical system by the following equation. 
\begin{equation}
\frac{\mathrm{d}\bm{x}}{\mathrm{d}t} =  \bm{f({x},{u}})
\label{eq:27}
\end{equation}
\noindent In the Trust Region optimization method, the nonlinear cost function $\added[id=new]{G}(\bm{\xi})$ is approximated by Taylor series for the $2^{nd}$ order and constrained on the so-called "trust region" where the approximation holds well enough. The constraint on the system is in the form of $||\bm{\xi - c}||_2 < \delta $, where $\bm{c}$ is the point around which the approximation is done. Then, the optimization problem in the Trust Region Reflective Newton method can be expressed as,
\begin{equation}
\begin{split}
\displaystyle{\min_{\bm{\xi}} \added[id=new]{G}(\bm{\xi})} = \displaystyle{\min_{\bm{\xi}}}
{\bm{(f(c)} + \nabla \bm{f^{T}(\xi-c)}} 
\\ {+ \frac{1}{2}(\bm{\xi-c})^{T}\bm{\Gamma}(\bm{\xi-c}))}
\end{split}
\label{eq:28}
\end{equation}
\noindent where, $\nabla$ is the gradient operator and $\Gamma$ is the Hessian matrix of $f(\bm{\xi})$. The solution for this problem is obtained by restricting the trust region sub-problem into a two-dimensional subspace as detailed in \cite{[49]}.  

\subsection{Model based on SINDYc}

Given the measured data and after discovering internal state data, the SINDYc algorithm can be used to fit a nonlinear sparse model as described in \cite{[44]}. Consider the system described by Equation \ref{eq:27}. The algorithm first develops a library of candidate functions $\Theta (\bm{x,u})$ which may also include essentially, nonlinear terms that are more likely to describe the underline dynamics of the system. If the functions are written in terms of the data matrices of state and control variables i.e $\bm{X} = [\bm{x}(t_1),\bm{x}(t_2),...\bm{x}(t_n)]$ and $\bm{U} = [\bm{u}(t_1),\bm{u}(t_2),...\bm{u}(t_n)]$, the candidate function library can be written as, 
\begin{equation}
{
\bm{\Theta^T} (\bm{X,U})
= \begin{bmatrix}
\bm{f_1}(\bm{X,U}) \\
\bm{f_2}(\bm{X,U})\\
\bm{f_3}(\bm{X,U})  \\
 .\\
 .\\
\end{bmatrix}}
\label{eq:35}
\end{equation}

\noindent  And then the system can be represented by the following equation, 
\begin{equation}
{\bm{\dot{X} = \Xi \Theta^T}  }
\label{eq:36}
\end{equation}
\noindent Where $\bm{\Xi}$ is the coefficient matrix. Usually, the coefficients are then determined using a sparse regression algorithm because most of the nonlinear systems do not have the same nonlinear term repeated in many state equations. A common sparse regression algorithm used in literature is LASSO \cite{[56]} i.e,
\begin{equation}
{\bm{\xi_k} = argmin_{\xi_k} ||\bm{\dot{X}}_k - \bm{\xi}_k \bm{\Theta}^T (\bm{X},\bm{U}) ||_2 + || \alpha \bm{\xi}_k||_1}
\label{eq:37}
\end{equation}
 \noindent  where $\bm{\xi_k}$ is the $k^{th}$ row of the coefficient matrix $\bm{\Xi}$ and $\bm{\dot{X}}_k$ is the $k^{th}$ row of the $\bm{\dot{X}}$ matrix. However, the method followed in this work is based on \added[id=new]{the algorithm} that is described in \cite{[44]}, where the resulting elements of $\bm{\Xi}$ after regression are hard thresholded one by one until it doesn't show a significant improvement in the model fit percentage (Equation \ref{eq:48}) for the validation dataset.

\section{{Methodology}}
    \label{Experiments}

\added[id=new]{Based on the theoretical background described in section \ref{Mathematical Modelling}, an experiment was set up to collect necessary data and a novel method was built to identify a data-driven model for BLDC motors that accounts for nonlinear friction. }

\subsection{Experimental Setup}

A BLDC motor (maxon international ltd.) with the specs given in Table \ref{tab:Motor_specs} with an inbuilt hall sensor system with the capability of measuring the angular displacement with a resolution of \added[id=new]{$5.45 ^0$ ($0.095$ rad)}  was used to build a setup to experiment the feasibility of using data-driven methods to identify nonlinear motor models. \added[id=new]{The angular velocity was calculated after averaging the position data using a moving average filter with a window size of 10 samples. The sampling time of all the signals was $12$ ms. Thus the resolution of angular velocity measurement was $0.095$ rad $/$ $(0.012 \times 10 )$ s = $0.79$ rad/s.} In addition, the setup was used to test the model-based friction compensation.

\begin{table}[tbp]
\renewcommand{\arraystretch}{0.5}
\setlength{\tabcolsep}{0.6em}
  \centering
  \caption{\small Motor Specifications}

  \vspace{1pt}
    \begin{tabular}{lccc}
    \toprule
    Specification & {     }& Value & Unit \\
    \midrule

    Nominal Voltage & {}&24 & V\\ [0.2cm]
    Nominal Current & {}&6.39 & A\\
    [0.2cm]
    Stall Current & {}&111 & A\\
    [0.2cm]
    Nominal Speed & {}&2720 & rpm\\
    [0.2cm]
    Max. Speed & {}&5000 & rpm\\
    [0.2cm]
    Nominal Torque & {}&457 & mNm\\
    [0.2cm]
    Stall Torque & {}&7910 & mNm\\
    [0.2cm]
    
    \bottomrule
    \end{tabular}%
    \label{tab:Motor_specs}%
    
\end{table}

\subsection{Excitation Signal}

\begin{figure}[tbp]
 \begin{center}
\includegraphics[width=0.5\textwidth]{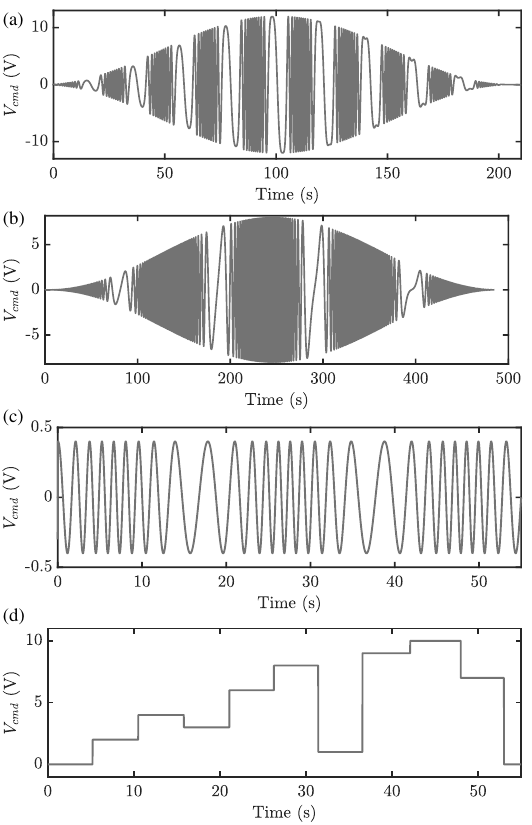}
 \vspace{-5pt}
 \caption{The excitation signals (commanded voltage) used for fitting and validation of motor model}
 \vspace{-25pt}
 \label{V_cmd}
 \end{center}
 \end{figure}

The excitation signals i.e. the commanded voltages were selected based on the required operating amplitude and frequency range of the motor for an active exoskeleton application. One signal was derived for model fitting and another three for validation. The model fitting signal depicted in Figure \ref{V_cmd} (a) has an amplitude range of $0-12$ V and a frequency range of $0.1 - 1 $ Hz. The first validation signal depicted in Figure \ref{V_cmd} (b) has a lower amplitude and a frequency range of $ 0.05 - 1 $ Hz. Furthermore, its frequency variation is slower. Therefore the signal (b) represents data from a different operating region of the motor in the sense of amplitude and frequency. To test the fitted model's performance in low amplitude the signal shown in Figure \ref{V_cmd} (c) was used, which has an amplitude range of $0 - 0.5$ V and a frequency range of $0.25 - 1$ Hz. Then the step signal with step amplitudes between $0-10$ V was used as shown in Figure \ref{V_cmd} (d). 

 \begin{figure}[t]
 \begin{center}
\includegraphics[width=0.5\textwidth]{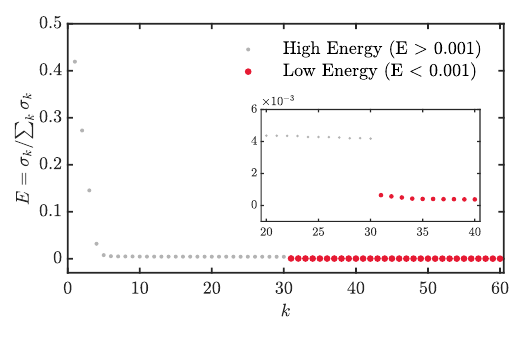}
 \vspace{-30pt}
 \caption{Singular value spectrum obtained after decomposing the time-delayed coordinates of the velocity data: low energy components selected to extract the hidden variable are shown in red}
 \vspace{-15pt}
 \label{fig:TDE_Eigen}
 \end{center}
 \end{figure}
 
\subsection{\added[id=new]{Internal State Extraction from TDE}}

\added[id=new]{It is noteworthy that this discovery becomes a major contribution to the research on motor friction modelling not only because of applying SINDYc on motor data but also using TDE to extract information about the internal state ($z$). Therefore first, the formulation of the method of extraction must be discussed.}

\added[id=new]{The motivation towards using delay embedding was the inability to build a closed-form model analytically or using data-driven techniques when one or more variables of the dynamical system are internal or unmeasurable. In delay embedding, a latent high-dimensional system is built using incomplete measurements by expanding the measured state variables into a higher-order set of states made out of the time history of the measured states \cite{[53]}.}

\added[id=new]{If $x(t)$ is the known, measured state in time $t$, its Hankel matrix, $\bm{H} \in \mathbb{R}^{n \times m}$ can be written as,}

\begin{equation}
\added[id=new]{
\bm{H} = \begin{bmatrix}
x(t_1) & x(t_2) & x(t_3) & ... & x(t_n) \\
x(t_2) & x(t_3) & x(t_4) & ... & x(t_{n+1}) \\
 .  &  .  &  .  &     &  . \\
 .  &  .  &  .  &     &  . \\
x(t_m) & x(t_{m+1}) & x(t_{m+2}) & ... & x(t_{m+n-1})\\
\end{bmatrix}
}
\label{eq:29}
\end{equation}
\noindent \added[id=new]{Where, $m$ is the number of time-delayed coordinates and $n$, is the number of measured time instances. The Hankel matrix is then subjected to economy SVD, to obtain the time series composition of Eigen modes.}
\begin{gather}
\added[id=new]{
\bm{H = U\Sigma V^{T}}} \label{eq:30}\\
\bm{\Sigma} = \begin{bmatrix}
\sigma_1 & 0 & 0& ... & 0 \\
0 & \sigma_2 & 0 & ... & 0 \\
 .  &  .  &  .  &     &  . \\
 .  &  .  &  .  &     &  . \\
0 & 0 & 0 & ... & \sigma_n\\
\end{bmatrix}
\label{eq:31}
\end{gather}
\noindent \added[id=new]{Here, $\bm{U} \in \mathbb{R}^{m \times n}$ is the left unitary matrix with Eigenvectors of the correlation matrix $\bm{HH^*}$ arranged in columns ($^*$ denotes the conjugate transpose). $\bm{\Sigma} \in \mathbb{R}^{n \times n}$ is a diagonal matrix with singular values ($\sigma_i$) arranged in descending order along the diagonal. $\bm{V} \in \mathbb{R}^{n \times n}$ is the right unitary matrix with Eigenvectors of the correlation matrix $\bm{H^*H}$ arranged in rows. The columns of the matrix $\bm{U}$ provide an orthonormal basis for the column space of $\bm{H}$, and the columns of $\bm{V}$ provide an orthonormal basis for the row space of $\bm{H}$ \cite{Brunton}. The matrix $\bm{U}$ encodes time-delayed patterns of velocity, and the matrix $\bm{V}$ encodes temporal patterns of velocity. The notion of the singular values here is that they represent how strongly the corresponding eigencomponent influences the dynamics of the time-delayed coordinates. Given the possibility that information about the asperity deformation is hidden inside velocity data, the authors hypothesised that this information must be represented by the lowest energy singular values. The basis behind this idea lies in two facts: first the magnitude of asperity deformation is of a tiny scale when compared with other states (displacement, velocity and current) and the second is that this variable is dynamic only in a short time range since the deformation gets saturated after reaching its maximum value according to the empirical findings of friction behaviour. }

\added[id=new]{Therefore, a method was formulated to rebuild the time-delayed coordinates after omitting prominent Eigen components by introducing a modified matrix $\bm{\hat{\Sigma}}$ which has a selected range of low energy eigenvalues, instead of $\bm{\Sigma}$. For example: $(n-l_0)$\textsuperscript{th} to $(n-l_0+l_1)$\textsuperscript{th}.}
\begin{equation}
\added[id=new]{
\bm{\hat{\Sigma}} = \begin{bmatrix}
0      & \cdots & 0      & 0      & \cdots & 0 & 0 & \cdots \\
\vdots &  \ddots & \vdots & \vdots & \ddots & \vdots& \vdots & \ddots \\
0      & \cdots & \sigma_{n-l_0} & 0 &\cdots & 0& 0& \cdots \\
0      & \cdots & 0 & \sigma_{n-l_0+1} & \cdots & 0& 0& \cdots \\
\vdots &  \ddots & \vdots & \vdots &\ddots& \vdots& \vdots& \ddots \\
0      & \cdots & 0 & 0 &\cdots & \sigma_{n-l_0+l_1}& 0& \cdots \\
0      & \cdots & 0 & 0 & \cdots & 0& 0& \cdots \\
\vdots  & \ddots & \vdots & \vdots & \ddots & \vdots & \vdots& \ddots \\
\end{bmatrix}
}
\label{eq:30.2}
\end{equation}
\noindent \added[id=new]{Now, the delay coordinates corresponding to low energy eigen components can be obtained as,}
\begin{equation}
\added[id=new]{
\bm{\hat{H} = U\hat{\Sigma} V^{T}}
}
\label{eq:31.2}
\end{equation}
\noindent \added[id=new]{Where, $\bm{\hat{H}}$ is the new Hankel matrix with low energy delay coordinates. The first coordinate of $\bm{\hat{H}}$ was directly used as a representation of the internal state.}  

\added[id=new]{The Hankel matrix $\bm{H}$ was built using the velocity data from the excitation signal shown in Figure \ref{V_cmd} \added[id=new]{(a)}. The internal state was then extracted using the method described in this section. After several trials, the order of delay embedding ($m$) was selected as $60$, since no significant improvement could be observed for internal state extraction in the sense of its resemblance to the values calculated using the discovered LuGre model. Figure \ref{fig:TDE_Eigen} shows the singular value spectrum of the delayed coordinates, where the selected low energy range to extract the internal state is from $k=31$ to $k=60$. The energy of singular values ($E$) is defined as the ratio between the corresponding singular value and the sum of all singular values ($\sigma_k / \Sigma_k \sigma_k$). The scientific reason behind this selection is the clear separation between the high energy singular values (defined as $E > 0.001$) and low energy singular values exactly at $30$\textsuperscript{th} value.}

\subsection{\added[id=new]{Model Discovery Using SINDYc}}

\added[id=new]{The generalized SINDy algorithm for systems with control discussed in Section \ref{Mathematical Modelling} was used to derive a nonlinear model for motor that accounts the friction. The candidate term library $\Theta(\textbf{X},\textbf{U})$ was composed  of position ($x$), velocity ($\dot{x}$), current ($I$) and commanded voltage ($U$). In addition, nonlinear functions: hyperbolic tangents of velocity and internal state ($\tanh  (a\dot{x})$ and $\tanh  (a\dot{z})$), and some other second and third-order products of velocity, internal state and commanded voltage were included. Here $a$ of the hyperbolic tangent function is a coefficient that ensures a close approximation of the unit step function.}
\added[id=new]{
\begin{multline}
\Theta (\bm{X},\bm{U})^\top
= 
    \left[\begin{matrix} \bm{X} \quad \bm{\dot{X}} \quad \bm{Z} \quad \bm{I} \quad \tanh (a \bm{\dot{X}}) \quad \tanh (a \bm{Z}) \end{matrix}\right. \\
    \left. \begin{matrix} \quad\qquad \bm{|\dot{X}|\dot{X}} \quad \bm{|\dot{X}|\dot{X^2}} \quad \bm{|Z|Z} \quad \bm{|Z|{Z^2}} \quad \bm{Z\dot{X}} \quad  \bm{Z^2\dot{X}}\end{matrix} \right. \\
    \left. \begin{matrix} \quad\qquad\qquad\qquad\qquad \bm{Z\dot{X}^2} \quad \bm{U} \quad \bm{ZU} \quad \bm{Z^2U} \quad \bm{ZU^2} \end{matrix} \right.]^\top \\ \hspace{-10pt}
\end{multline}}
\added[id=new]{The hyperbolic tangent functions of velocity and internal state ($\tanh(a\dot{x})$ and $\tanh(az)$) were included in the candidate term library since they closely approximate the Coulomb friction, which behaves like a step function in velocity. The second and third-order polynomial terms of velocity ($|\dot{x}|\dot{x}$ and $|\dot{x}|\dot{x}^2$) were included since the viscous friction can be more accurately represented by a higher-order polynomial of velocity instead of a first-order polynomial. The frictional force consists of a component which is a function of the internal state which represents the asperity deformation. This could be characterised by a general linear stress-strain relationship by considering only the first-order term of $z$. However, authors assumed that better characterisation could be applied by considering the dependency of asperity stress on the higher order terms: $|z|z$ and $|z|z^2$. The cross terms of $z$ and $\dot{x}$, and $z$ and $u$ were included to model the small undetectable errors that occur in position and current measurements due to asperity deformation.}

\added[id=new]{The sparse coefficient matrix $\bm{\Xi}$ \added[id=new]{(Equation \ref{eq:36})} was then obtained by the SINDYc algorithm. After iteratively omitting the insignificant coefficients, it resulted in a sparse matrix. The four-state equations can be written as a linear combination of the terms in the candidate term library (Equations \ref{eq:43} - \ref{eq:46}). Note that the coefficient $a_1$ is theoretically equal to one. However, it was included in the coefficient matrix to verify that the time derivative of the position has been calculated correctly. }
\begin{equation}
\begin{split}
\added[id=new]{
\mathit{\frac{\mathrm{d}x}{\mathrm{d}t} = 
a_1\dot{x} }
}
\end{split}
\label{eq:43}
\end{equation}
\vspace{-30pt}

\begin{equation}
\begin{split}
\added[id=new]{
\frac{\mathrm{d}\dot{x}}{\mathrm{d}t} = b_1\dot{x} + b_2z + b_3I + b_4\tanh(a\dot{x}) + b_5|\dot{x}|\dot{x}+ b_6|\dot{x}|\dot{x}^2} \\ 
\added[id=new]{+ b_7|z|z +  b_8|z|z^2 
 + b_9z\dot{x} + b_{10}z^2\dot{x}+ b_{11}z\dot{x}^2 + b_{12}u} \\
 \added[id=new]{+ b_{13}zu + b_{14}zu^2
 }
\end{split}
\label{eq:44}
\end{equation}
\vspace{-50pt}

\begin{equation}
\begin{split}
\added[id=new]{
\frac{\mathrm{d}z}{\mathrm{d}t} = c_1\dot{x} + c_2z + c_3\tanh(a\dot{x}) + c_4\tanh(az)  +  c_5|\dot{x}|\dot{x} }\\
\added[id=new]{+  c_6|z|z^2 }
\end{split}
\label{eq:45}
\end{equation}
\vspace{-40pt}

\begin{equation}
\begin{split}
\added[id=new]{
\frac{\mathrm{d}I}{\mathrm{d}t} = d_1\dot{x} + d_2z + d_3I + d_4\tanh(a\dot{x}) + d_5\tanh(az)  + }\\ 
\added[id=new]{d_6|\dot{x}|\dot{x} +  d_7|z|z^2 + d_8z^2\dot{x} + d_9z\dot{x}^2  + d_{10}u + d_{11}z^2u 
}
\end{split}
\label{eq:46}
\end{equation}
\vspace{-40pt}

\begin{equation}
\begin{split}
\added[id=new]{
F_f = J(b_1\dot{x} + b_2z + b_4\tanh(a\dot{x}) + b_5|\dot{x}|\dot{x} + b_6|\dot{x}|\dot{x}^2 }\\ \added[id=new]{+ b_7|z|z +  b_8|z|z^2 + b_9z\dot{x} + b_{10}z^2\dot{x} + b_{11}z\dot{x}^2)
}
\end{split} 
\label{eq:47}
\end{equation}

\added[id=new]{Equation \ref{eq:44} suggests that velocity dynamics depend nonlinearly on velocity, internal state and also the commanded voltage. The dependency on commanded voltage can be interpreted as compensation for the current measurement error. Therefore, the last three terms of the equation can be considered complementary terms for the current dependency ($b_3I$). The remaining terms can be interpreted as components of nonlinear friction. Therefore the newly identified friction function can be written as Equation \ref{eq:47}. The dynamics of the internal state depend nonlinearly on velocity and the internal state itself. This result aligns with the empirical results discussed in the literature. The dependency on $\tanh$ terms can be interpreted as the Coloumb friction. Lastly, the current dynamics depend on the current linearly, like in the linear model but depend on both velocity and internal state nonlinearly. Furthermore, the dependency on commanded voltage is complicated by the influence of the internal state. The complicated dependency of current dynamics on the internal state can be interpreted as compensation for the measurement errors of position, velocity and current at the pre-sliding friction regime due to the low resolution of the measurements. }

 \begin{figure}[t]
 \begin{center}
\includegraphics[width=0.5\textwidth]{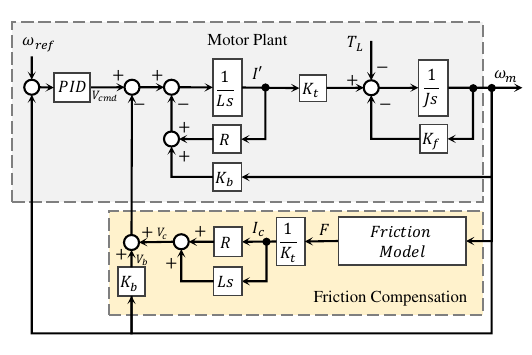}
 \vspace{-20pt}
 \caption{Block diagram of the motor speed control algorithm with friction compensation}
 \vspace{-5pt}
 \label{fig:Block}
 \end{center}
 \end{figure}
 
\subsection{Friction Compensation}

A friction compensation algorithm based on the SINDYc model was built to test the effectiveness of the newly identified model in friction compensation applications. The algorithm was based on the feedback compensation strategy that is theorized in this section. The governing equations of the standard linear motor model can be expressed as follows:
\begin{gather} 
    {J\dot{\omega}_m} = {K_t}I-{\added[id=new]{F_f}} \label{eq:1}\\
    {L\dot{I}} = -RI-{K_b}{\omega_m} +{V_{cmd}} \label{eq:2}
\end{gather}

\noindent Here, 
$J$ is the inertia of the rotor, $\omega_m$ is the motor speed, $I$ is the motor current, \added[id=new]{$F_f$} is the frictional torque, $V_{cmd}$ is the commanded voltage and,  $R$ and $L$ are the resistance and the inductance of the rotor winding. Then, $K_t$, $K_b$ represents the torque and back-EMF constants. The model described by the Equations \eqref{eq:1} and \eqref{eq:2} may incorporate nonlinearities in the friction term (\added[id=new]{$F_f$}) making it a nonlinear model. 

Now suppose that the current needed to compensate for the friction force is $I_{c}$. Then,
\begin{equation}
{{K_t}I_{c} = \added[id=new]{F_f} }
\label{eq:38}
\end{equation}

\noindent Furthermore, if the compensation voltage necessary to generate this current is $V_{c}$, then adding this to Equation \ref{eq:2} and modifying the new current as $I' = I + I_{c}$ gives,
\begin{equation}
{L(\dot{I'} )} = -R(I')-{K_b}{\omega_m} +{V'_{cmd}} + V_{c} 
\label{eq:39}
\end{equation}

\noindent Rearranging gives an expression for compensation voltage that must be applied to compensate for the frictional forces,
\begin{equation}
{V_{c}  = \dot{I}_{c} + RI_{c} } 
\label{eq:40}
\end{equation}

\noindent Using Equations \ref{eq:39} and \ref{eq:40}, the commanded voltage can be modified so that the system behaves like a frictionless system. A block diagram representation of the friction compensation control algorithm is illustrated in Figure \ref{fig:Block}.

\section{Results}
    \label{Results}
This section presents the results of the model identification and friction compensation schemes tested during the experiments on the BLDC motor described in the previous sections. First, the identification of LuGre model parameters using a nonlinear grey box model estimation technique is presented. It is followed by the internal state extraction, SINDYc model identification results and friction compensation.

\subsection{LuGre Model \added[id=new]{Parameters}}

\begin{figure}[tbp]
 \begin{center}
\includegraphics[width=0.5\textwidth]{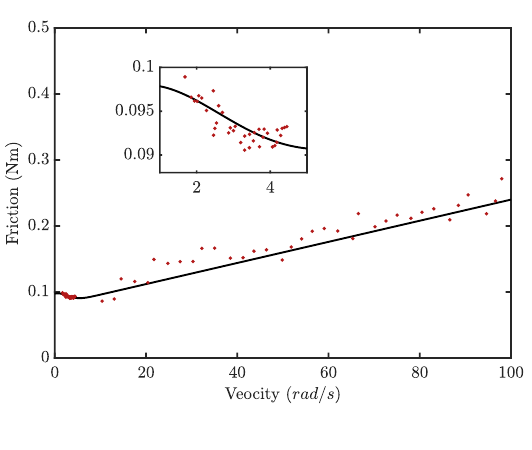}
 \vspace{-30pt}
 \caption{Velocity vs friction data obtained from steady-state velocity experiments with fitted model}
 \vspace{-15pt}
 \label{Lugre}
 \end{center}
 \end{figure}

Constant voltage signals ranging from $0.4$ V  to $0.6$ V were given to the motor to measure the steady-state velocities. Using this data, the static parameters of the LuGre model were calculated. Later, slow voltage ramp signals with gradients ranging from $9.5$  to $12$ ${mVs^{-1}}$ were given and the current and velocity were measured. The measured data was used to derive an initial approximation for dynamic parameters. Then nonlinear grey box model estimation technique was used to estimate more accurate values for dynamic parameters. A comparison between the measured friction value and the value predicted by the fitted model is shown in Figure \ref{Lugre}. The estimated parameters are presented in the Table \ref{tab:Lugre_parameters}. 


 \begin{table}[tbp]
\renewcommand{\arraystretch}{0.5}
\setlength{\tabcolsep}{0.6em}
  \centering
  \caption{\small Estimated LuGre Parameters}

  \vspace{1pt}
    \begin{tabular}{cccccc}
    \toprule
    $\alpha_0$ & $\alpha_1$ & $\alpha_2$ & $v_0$ & $\sigma_0$& $\sigma_1$ \\
    \midrule
    0.0800 & 0.0175  & 0.0016 & 3.6760 & 317.2250 & 22.2464 \\
    \bottomrule
    \end{tabular}%
\label{tab:Lugre_parameters}%
  \vspace{-10pt}
\end{table}

\subsection{\added[id=new]{Internal State} Extraction from  \added[id=new]{TDE}} 

\added[id=new]{The extracted internal state data using the method described in Section 4.3 was compared with the data obtained from the discovered LuGre model.} The comparison is shown in Figure \ref{fig:Results_1}. It must be noted that the values graphed here are normalized. In the visual comparison of the values extracted from TDE and values from the LuGre model, a very good qualitative correlation can be observed, specifically, at low amplitudes (first and last $50$ seconds on the graph). However, the correlation is lower at a high amplitude range (from $50$ s to $150$ s), for which a possible reason is the low signal-to-noise ratio (SNR) at high amplitudes. It must also be mentioned that the scale of original values obtained from TDE was scaled up by around $\times 50$ when compared with the values from the LuGre model after normalizing. The authors did not consider this as an issue since the $z$ variable will be incorporated with appropriate coefficients after fitting the model using SINDYc. 

\subsection{SINDYc Model}

\vspace{0pt}

\begin{figure}[tbp]
\begin{center}
\includegraphics[width=0.5\textwidth]{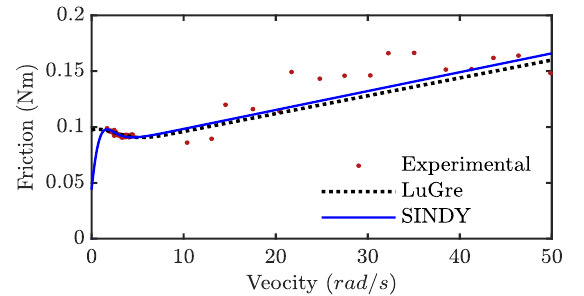}
 \vspace{-13pt}
 \caption{Comparison of speed vs friction torque characteristics of LuGre and SINDY models.}
 \vspace{-20pt}
 \label{fig:F_V}
 \end{center}
 \end{figure}

 \begin{figure}[tbp]
\begin{center}
\includegraphics[width=0.5\textwidth]{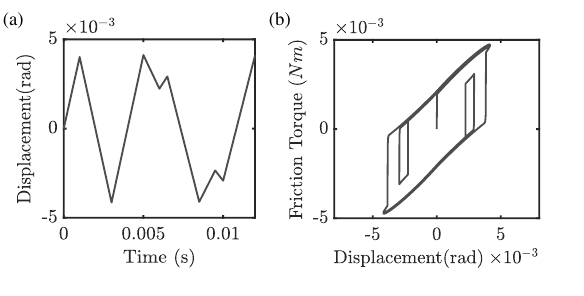}
 \vspace{-20pt}
 \caption{The pre-sliding hysteresis behaviour with non-local memory characterised by the new friction model: (a) position signal as a function of time (b) friction variation with pre-sliding displacement}
 \vspace{-20pt}
 \label{fig:Hys}
 \end{center}
 \end{figure}

  \begin{figure*}[t]
 \begin{center}
\includegraphics[width=1\textwidth]{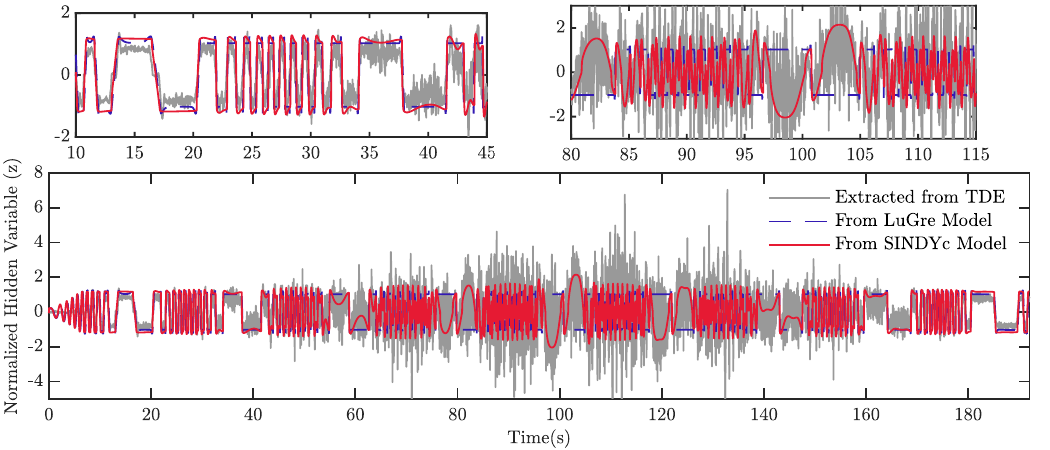}
 \vspace{-10pt}
 \caption{Comparison of normalized internal state, $z$ (asperity deformation) obtained using SVD of time-delayed coordinates, LuGre model and SINDYc model}
 \vspace{-15pt}
 \label{fig:Results_1}
 \end{center}
 \end{figure*}

  \begin{figure*}[ht!]
 \begin{center}
\includegraphics[width=1\textwidth]{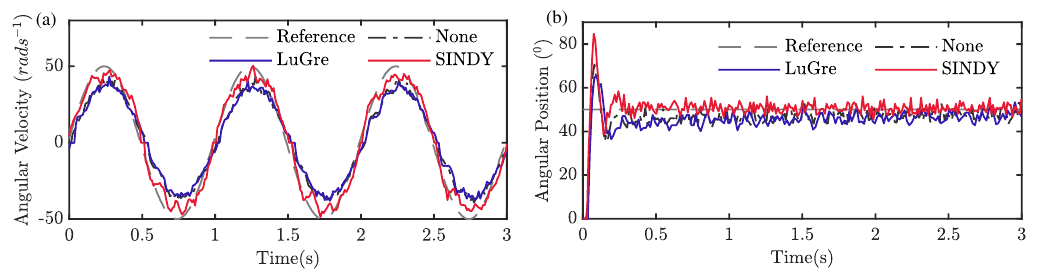}
 \vspace{-15pt}
 \caption{Comparison of reference velocity tracking performances of LuGre and SINDYc models}
 \vspace{-15pt}
 \label{fig:Friction_Comp}
 \end{center}
 \end{figure*}

\begin{table*}[t!]
\centering
\caption{Fit Percentages}
    \begin{tabular}{ccccccccc}
    \toprule
    \multicolumn{1}{c}{\multirow{3}[4]{*}{Model}} & \multicolumn{8}{c}{Model fit percentage ($\%$)} \\
    \cmidrule(l){2-9} 
    {}& \multicolumn{2}{c}{(a)}  & \multicolumn{2}{c}{(b)} &\multicolumn{2}{c}{(c)} &\multicolumn{2}{c}{(d)} \\
    \cmidrule(l){2-9}     
    {} & $x$ & $\dot{x} $ & $x$ & $\dot{x}$ & $x$ & $\dot{x}$ & $x$ & $\dot{x}$ \\
    \cmidrule(r){1-1} \cmidrule(r){2-3} \cmidrule(r){4-5}
    \cmidrule(r){6-7}
    \cmidrule(r){8-9}
    Linear  & 78.43  & 73.76 & 83.89   & 75.07  & -3.71  & 16.68  & 95.78  & 86.89   \\
    LuGre  & 74.30	& 77.83	& 85.35	& 75.68	& -9.14	& -11.93	& 94.64	& 87.94 \\
    SINDYc   &91.32	&91.46	&87.45&	92.67	&45.58&	41.87	&96.03&	90.82 \\
    \bottomrule
    \end{tabular} \\
    (a),(b),(c) and (d) denote the excitation signals shown in Figure \ref{V_cmd}.
\label{tab:Results_comparison}
    \vspace{-5pt}
\end{table*}%

The internal state estimated using the discovered SINDYc model has been plotted in Figure \ref{fig:Results_1} together with the values obtained from the LuGre model and values extracted using TDE. The speed vs torque characteristics of the SINDYc model were revealed by running a simulation using the Equations \ref{eq:44}-\ref{eq:47}. Steady-state velocity was determined for constant voltage commands. Figure \ref{fig:F_V} shows the results which also include the experimental data and results obtained using the LuGre model. To check the pre-sliding behaviour of the SINDYc friction model, a simulation was run to characterise the relationship between displacement and friction torque. The results are shown in Figure \ref{fig:Hys}, and it is evident that the new SINDYc model features the hysteresis with non-local memory as proved by empirical results. However, any experimental results to compare the simulation results could not be obtained due to the low resolution of the sensors. However, the authors believe that the qualitative similarity of the behaviour of the newly identified friction model to the empirical results presented in the literature is sufficient to accept it as a valid model.

\subsection{Friction Compensation}
 \begin{figure*}[t]
 \begin{center}
\includegraphics[width=0.97\textwidth]{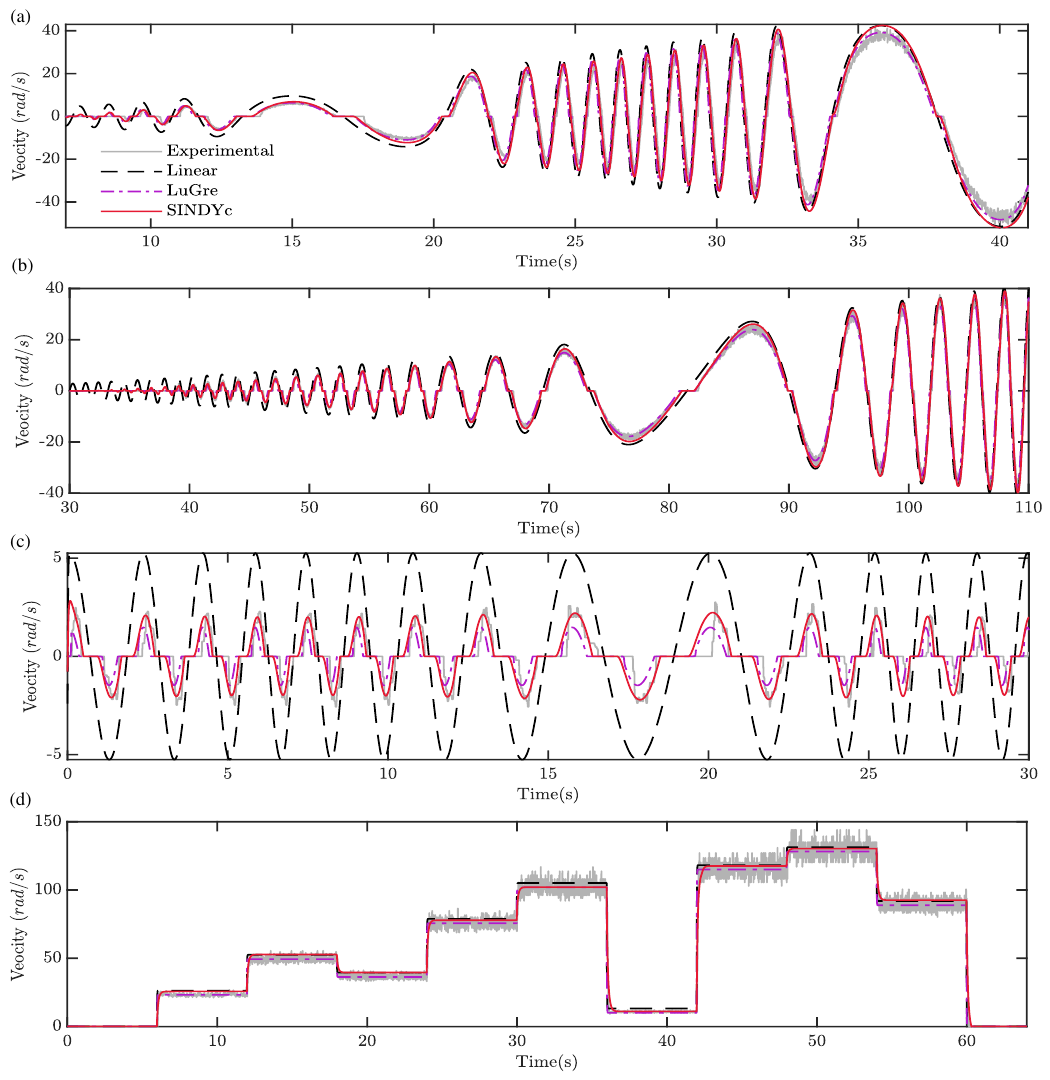}
 \vspace{-5pt}
 \caption{Comparison of the velocity predicted by the three models: Linear, LuGre and SINDYc with the experimental data}
 \vspace{-15pt}
 \label{fig:Results_2}
 \end{center}
 \end{figure*}
The friction compensation algorithm described in Section \ref{Experiments} was implemented on the BLDC motor. The gain of the controller was set so that it could ensure that the system would not become unstable due to changes that happened in model parameters during the operation. Two input cases, first a sinusoidal reference velocity signal and second a step signal were used to compare how well the velocity tracking happens in systems with and without friction compensation. The results are shown in Figure \ref{fig:Friction_Comp}. The graphs also give an insight into the superiority of the newly discovered model over the LuGre model. 
\subsection{Comparison of Results}
The performance of models in predicting motor states was measured using a goodness of fit function based on normalized root mean square error (NMSE) described by Equation \ref{eq:48} below.
\begin{equation}
\mathit{fit  =  100 \Bigl( 1 - \frac{||y - \hat{y} ||}{||y - \Bar{y} ||} \Bigl) }
\label{eq:48}
\end{equation}

\noindent Here, $y$ is the measured value, $\hat{y}$ is the estimated value and $\Bar{y}$ is the mean value of $y$. The fit percentage values of all four models for all four excitation signals are shown in Table \ref{tab:Results_comparison}. It is noteworthy that except for the signal (c), i.e. low-velocity operation, in all other scenarios, all three models are performing at least fairly well. Out of those, the SINDYc model outperforms the other three models. For the low-velocity scenario, SINDYc outperforms all other models showing outstanding performance when compared with linear and LuGre models. \added[id=new]{The low accuracy in estimation at low velocity is mainly due to the error in the experimental data that resulted from the low resolution of the angular velocity measurement which is only $0.79$ rad as described in Section \ref{Experiments}}. To summarize, the data-driven model based on SINDYc was able to outperform the conventional linear and nonlinear models for all the cases considered. 
 
A comparison between measured velocity and predicted velocity values from the models for the fit dataset is shown in Figure \ref{fig:Results_2}(a). Furthermore, the Figure \ref{fig:Results_2}(b) shows the responses for the first validation signal. Here, only a selected epoch of the time series data is depicted for clarity. It is observable that at high velocities, the LuGre model performs slightly better than the data-driven model, but at low-velocity values, the SINDYc-based model takes over. Thus to clarify this observation it is important to observe the low-velocity behaviour using an excitation signal that has a low amplitude range. For this purpose, the response for the excitation signal shown in Figure \ref{V_cmd}(c) has been presented in Figure \ref{fig:Results_2}(c). From this graph, it is evident that LuGre model always predicts a lower value for velocity when compared with experimental values. On the other hand, the SINDYc model predicts closer values. Figure \ref{fig:Results_2}(d) shows the responses for step signal, and it can be seen that the SINDYc model outperforms the other two models.

\section{Conclusions}
    \label{Conclusion}
\vspace{5pt}
This article proposes a novel nonlinear model for predicting the states of an electric motor including friction. The model was developed by applying the SINDYc algorithm to velocity, internal state and current data of a BLDC motor, which resulted in a sparse coefficient matrix for a model with $13$ nonlinear terms. Furthermore, the internal state variable of friction i.e. asperity deformation was extracted using delay embedded coordinates of motor velocity using SVD and eliminating high energy singular values. 

The discovered nonlinear dynamical model was compared with the conventional linear model and a model that incorporates the LuGre friction model. The new model outperformed the other two models showing superiority in predicting motor states accurately. Furthermore, the TDE-based method was proven to be successful in extracting the internal state of friction. The developed model was then used in a feedback friction compensation algorithm to estimate friction and use that information to produce a control signal that accounts for the friction. Promising results could be obtained when compared with the LuGre model in compensating the friction.

The authors intend to conduct additional experiments utilising the identified model and the corresponding friction compensation algorithm to assess its robustness against external disturbances, such as fluctuating load conditions. Additional research can be conducted to develop an adaptive friction compensation algorithm that considers the changing load, motor orientation, and temperature conditions using the model discovery methodology established in this study.  
    
\section*{Acknowledgements}
    \label{Acknowledgements}

The authors acknowledge the support given to carry out the experiments by the Bionics Laboratory of the Department of Mechanical Engineering, University of Moratuwa. 
    
\section*{Statements and Declarations}
    \label{Statements}
\small
\textbf{Funding} The authors declare that no funds, grants, or other support were received for conducting this study.\\

\noindent\textbf{Competing Interests} The authors have no relevant financial or non-financial interests to disclose.\\

\noindent\textbf{Conflict of Interest} The authors declare that they have no conflict of interest.\\

\noindent\textbf{Author Contributions} All authors contributed to the study conception and design. Material preparation, data collection and analysis were performed by both authors. The first draft of the manuscript was written by Nimantha Dasanayake and Shehara Perera commented on and made modifications in subsequent versions. All authors read and approved the final manuscript.\\

\noindent\textbf{Data Availability} The datasets generated during and analysed during the current study are not publicly available due to the intellectual property rights policy of the University of Moratuwa. But are available from the corresponding author on reasonable request. 
    
\bibliographystyle{spmpsci}      
    \bibliography{sn-bibliography}

\end{document}